\documentclass[amsmath,natbib,rotating]{mn2e}
\usepackage{graphicx}
\title[Complex organic molecules in cold gas]{Formation of complex
  organic molecules in cold objects: the role of gas phase reactions}
\author[Balucani et al.]{Nadia Balucani$^{1,2,3}$\thanks{E-mail: nadia.balucani@unipg.it}, 
Cecilia Ceccarelli$^{2,3}$\footnotemark[1]\thanks{E-mail: Cecilia.Ceccarelli@obs.ujf-grenoble.fr},
Vianney Taquet$^{4}$\footnotemark[2]\thanks{E-mail: vianney.taquet@nasa.gov}\\
$^{1}$Universit\`a di Perugia, Dip. di Chimica, Biologia e
   Biotecnologie, I-06123 Perugia, Italy\\
$^{2}$ Univ. Grenoble Alpes, IPAG, F-38000 Grenoble, France \\
$^{3}$ CNRS, IPAG, F-38000 Grenoble, France\\
$^{4}$ Astrochemistry Laboratory and The Goddard Center for Astrobiology, Mailstop 691,\\ NASA Goddard Space Flight Center, 8800 Greenbelt Road, Greenbelt, MD 20770, USA\\
}

\begin{document}

    \date{Received - ; accepted -}

\pagerange{\pageref{firstpage}--\pageref{lastpage}} \pubyear{2014}

\maketitle

\label{firstpage}

\begin{abstract}
  While astrochemical models are successful in reproducing many of the
  observed interstellar species, they have been struggling to explain
  the observed abundances of complex organic molecules. Current models
  tend to privilege grain surface over gas phase chemistry in their
  formation. One key assumption of those models is that radicals
  trapped in the grain mantles gain mobility and react on lukewarm
  ($\ga30$ K) dust grains. Thus, the recent detections of methyl
  formate (MF) and dimethyl ether (DME) in cold objects represent a
  challenge and may clarify the respective role of grain surface and
  gas phase chemistry. We propose here a new model to form DME and MF
  with gas phase reactions in cold environments, where DME is the
  precursor of MF via an efficient reaction overlooked by previous
  models. Furthermore, methoxy, a precursor of DME, is also
  synthetized in the gas phase from methanol, which is desorbed by a
  non-thermal process from the ices. Our new modelreproduces fairy
  well the observations towards L1544. It also explains, in a natural
  way, the observed correlation between DME and MF. We conclude that
  gas phase reactions are major actors in the formation of MF, DME and
  methoxy in cold gas. This challenges the exclusive role of
  grain-surface chemistry and favours a combined grain-gas chemistry.
 \end{abstract}


\section{Introduction}\label{sec:introduction}
Relatively complex (i.e. containing more than six atoms) organic
molecules, referred in the literature as COMs, were first discovered
in the warm ($\ga 100$ K) and dense ($\ga 10^6$ cm$^{-3}$) regions
surrounding high mass protostars, called hot cores (Blake et
al. 1987). Several models were then developed to explain the observed
abundances of those COMs (Millar et al. 1991; Charnley et al. 1992;
Caselli et al. 1993). The basic idea of those models was that simple
hydrogenated molecules like H$_2$CO, CH$_3$OH and NH$_3$ are formed
during the cold pre-collapse phase on the cold ($\la 20$ K) grain
surfaces (e.g. Tielens \& Hagen 1980). Upon heating from the accreting
central protostar, the grain ices sublimate injecting those molecules
in the gas-phase and starting in this way what was then called a warm
gas-phase chemistry. This paradigm has lasted for almost two decades,
until new observations (e.g. Cazaux et al. 2003; Bottinelli et
al. 2007; Requena-Torres et al. 2006), laboratory experiments (Geppert
et al. 2007) and theoretical studies (Horn et al 2004) challenged
it. The warm gas phase chemistry paradigm left the place to the
grain surface chemistry.

In the new paradigm, grain surface chemistry is not only responsible
for the hydrogenated molecules of the pre-collapse phase, but also for
the (almost) whole set of observed COMs. The basic idea of this class
of models (Garrod \& Herbst 2006) is that radicals trapped in the iced
mantles acquire mobility and react forming COMs when the dust
temperature reaches $\sim30$ K. The origin and quantity of the trapped
radicals depends on the adopted model. They might be the pieces of
iced-species (e.g. methanol) broken by the FUV photons produced by the
interaction of cosmic rays with H$_2$ (e.g. Garrod \& Herbst 2006;
Chang \& Herbst 2014), or, alternatively, they are the result of the
incomplete hydrogenation of the simple mother species (e.g. Taquet et
al. 2012). Recent observations, however, show the presence of some
COMs -notably methoxy (CH$_3$O), methyl formate (HCOOCH$_3$, MF) and
dimethyl ether (CH$_3$OCH$_3$, DME)- in regions where the dust
temperature is less than 30 K: pre-stellar cores (Bacmann et al. 2012;
Vastel et al. 2014) and cold envelopes of low mass protostars (Oberg
et al. 2010; Cernicharo et al. 2012; Jaber et al. 2014). Two recent
models have been published trying to explain these new observations,
assuming that the detected COMs are present inside the cold and dense
core (Vasyunin \& Herbst 2013, hereinafter VH2013; Chang \& Herbst
2014).

More recent observations towards the pre-stellar core L1544 (Bizzocchi
et al. 2014; Vastel et al. 2014) added a crucial information: the
methanol emission originates from the outer shell of the cold core,
where the H$_2$ density is $\sim 3\times10^4$ cm$^{-3}$ and the
temperature is $\sim10$ K, and not from the inner and denser region,
as previously assumed. Therefore, methanol is ejected from the grain
mantles into the gas because of some non-thermal ice desorption, as it
was previously claimed for water (Caselli et al. 2012). Finally,
Vastel et al. (2014) argued that the COMs emission detected in L1544
also comes from the same outer shell and speculated that this is
likely the case also for the other cold sources where COMs have been
detected (Bacmann et al. 2012; Oberg et al. 2010; Cernicharo et
al. 2012).

Following these last findings and building up on the VH2013 model, we
propose here a new model where gas phase reactions, triggered by the
non-thermal desorption of methanol from the ices, play a major role in
the formation of methoxy, MF and DME in cold regions.  Specifically,
we consider a set of gas phase reactions, which were previously
overlooked or whose coefficients we have refined based on recent experimental
and theoretical works.
\begin{figure}
  \includegraphics[width=9cm,angle=180]{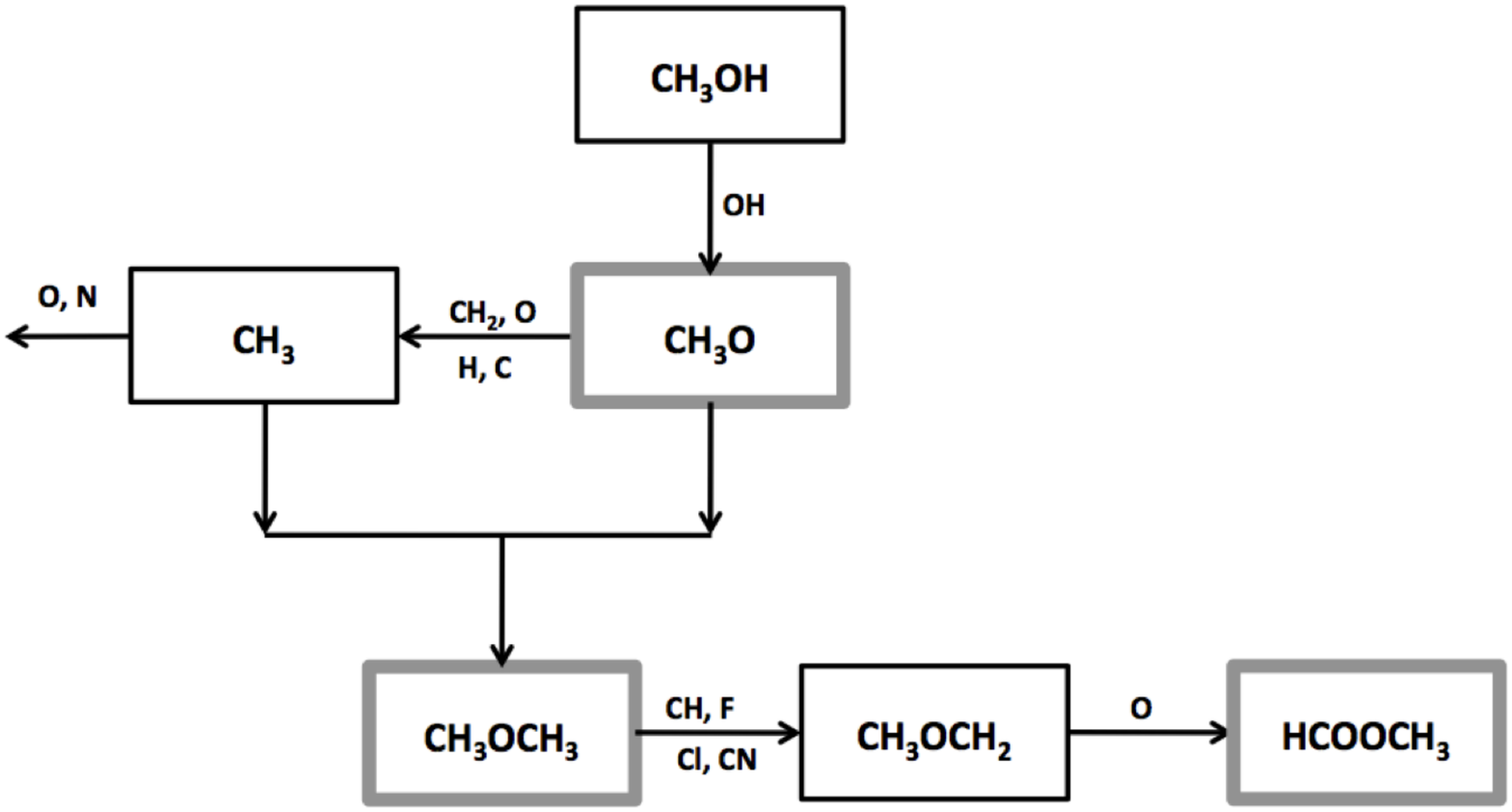}
  \caption{Scheme of the added reactions that form DME, MF and
    methoxy (boxes with grey large contours).}
  \label{abundances}
\end{figure}

\section{Model and adopted chemical network}\label{sec:chem-netw-model}
\noindent
\subsection{The model}
We used the GRAINOBLE code (Taquet et al. 2012, 2014) to compute the
mantle species from hydrogenation of CO and O. Briefly, GRAINOBLE
couples the gas-phase and grain-surface chemistry with the rate
equations approach introduced by Hasegawa et al. (1992), and takes
into account the multilayer and porous nature of the ice
mantles. Species accrete on the grain surfaces and are desorbed by
cosmic ray-induced heating of grains (Hasegawa \& Herbst 1993) and by
a non-thermal process that injects into the gas phase 1\% of the
species formed on the surface, following a process called ``chemical
desorption'' (Garrod et al. 2007).  We followed the chemical
composition evolution of a gas with a H density of $6\times 10^4$
cm$^{-3}$ and a temperature of 10 K (to simulate the conditions in
L1544; Vastel et al. 2014). We started with a gas having the initial
abundances quoted by Wakelam \& Herbst (2008) for the diffuse clouds
(their Table 1, column EA2; note that hydrogen is initially
molecular).  We follow the chemical evolution for $10^6$ yr.

In addition to this ``standard'' model, we run a few more models to
evaluate the impact on the results when some crucial parameters are
changed. Specifically, we considered the following cases: (i) a 0\% or
10\% non-thermal desorption fraction, the latter chosen based on the
experiment by Dulieu et al. (2013), although new experiments by
Minissale \& Dulieu (2014) suggest a dependence of this value on the
substrate and other parameters; (ii) a gas with a 5 times larger
density, and with 20 K temperature; (iii) a high and low Cl and F
abundances to evaluate their impact on the results, as some important
reactions introduced in this work depend on them (see next \S); (iv)
the rate coefficient of reaction (2) of Table 1 reduced by a factor 10
(see below).

\noindent
\subsection{The chemical network}\label{sec:chemical-network}
The surface chemical network is based on the one presented by Taquet
et al. (2013) but does not include the formation of deuterated
species. We used the binding energies listed in Taquet et al. (2014)
and the energy barrier for the CO and H$_2$CO hydrogenation recently
computed by Rimola et al. (2014).

\begin{table*}
   \centering
   \begin{tabular}{ l | c | l } 
     Reaction & Rate coefficient & References  \& \\
                 &  (cm$^3$ s$^{-1}$ at 10 K)  &  Notes \\ \hline
(1) OH + CH$_3$OH $\rightarrow$ CH$_3$O + H$_2$O                & 3.0$\times10^{-10}$  &  1\\
(2) CH$_3$O + CH$_3$ $\rightarrow$ CH$_3$OCH$_3$ + photon  & 3.0$\times10^{-10}$ & see text \\
(3) O + CH$_3$OCH$_2$  $\rightarrow$ HCOOCH$_3$ + H            &2.0$\times10^{-10}$ & 2, 3\\
(4) F + CH$_3$OCH$_3$  $\rightarrow$ CH$_3$OCH$_2$ + HF      & 2.0$\times10^{-10}$ & 2 \\
(5) Cl + CH$_3$OCH$_3$  $\rightarrow$ CH$_3$OCH$_2$ + HCl    &
2.0$\times10^{-10}$ & 4 \\  \hline
   \end{tabular}
   \caption{List of the added reactions that form DME, MF and
     methoxy.  References: 1- Shannon et al. 2013; 2- Hoyermann \&
     Nacke 1996; 3-  Song et al. 2005; 4- Wallington et al. 1988.}
   \label{tab:ini-abu}
 \end{table*}
 For the gas phase chemistry, we used the OSU2009 network (Harada \&
 Herbst 2008; http://faculty.virginia.edu/ericherb) as a basis, and
 added a set of new reactions for the formation of MF, DME and
 methoxy, and involving atomic chlorine and fluorine for the
 conversion of DME into its radical CH$_3$OCH$_2$. The scheme of the
 chemical reactions added in this work is shown in Fig. 1 and their
 list is reported in Table 1. We comment the novelties of the adopted
 network, for each species, in the following.

 \noindent \underline{Methoxy} is mostly formed by reaction 1.  This
 reaction was also considered by VH2013, with a rate coefficient equal
 to $4\times 10^{-11}$ cm$^3$s$^{-1}$. We instead adopted the value
 calculated by Shannon et al. (2013) at 20 K, $3\times 10^{-10}$
 cm$^3$s$^{-1}$, which, according to their suggestion, is likely a
 lower limit.  Note that the reactions involving the methoxy reported
 by VH2013 in Table 1 are not in line with the cited NIST Chemical
 Kinetics Database ({\it
   http://kinetics.nist.gov/kinetics/Search.jsp}), where the rate coefficients
  are different from those quoted by VH2013\footnote{The
   rate coefficients reported by VH2013 for the reactions CH$_3$O + H
   and CH$_3$O + O are erroneous: the NIST values are
   $3\times10^{-11}$ and $2.5\times 10^{-11}$ cm$^3$s$^{-1}$,
   respectively, rather than $1.6\times10^{-10}$ and
   $1.0\times10^{-10}$. In addition, the dominant reaction channel for
   CH$_3$O + H is the one leading to H$_2$CO + H$_2$ and not the one
   leading to CH$_3$ + OH, as claimed by VH2013.}.

 \noindent \underline{DME} is mainly formed by reaction 2. The rate
 coefficient for the radiative association of methyl and methoxy is
 not kwown. Recent theoretical work by Vuitton et al. (2012) suggests
 that radiative association can be very efficient when two relatively
 large radicals interact and form a stable molecule, if two-body
 exothermic channels are not available to the system. VH2013 quote the
 presence of a competitive exothermic channel leading to H$_2$CO +
 CH$_4$ and, therefore, employed a rate coefficient at 10 K of
 $2.6\times10^{-11}$ cm$^3$s$^{-1}$. However, a recent study by
 Sivaramakrishnan et al. (2011) has clearly pointed out that the
 reaction CH$_3$O + CH$_3$ $\rightarrow$ H$_2$CO + CH$_4$ is a direct
 abstraction process, not correlating with the bound intermediate
 CH$_3$OCH$_3$.  For this reason, we have increased the rate
 coefficient for reaction 2 by an order of magnitude, as there are no
 alternative two-body reaction channels and the hot DME formed after
 the association of CH$_3$ and CH$_3$O can only dissociate back or
 stabilize by radiative association. Yet, we have run also a model
 with the reaction 2 rate coefficient a factor ten smaller to evaluate the impact
 on the results.

 \noindent \underline{MF} is formed by reaction 3, namely the oxidation of
 CH$_3$OCH$_2$. This radical is formed by several reactions involving
 DME (Fig. 1) and common atomic and radical species, such as atomic
 chlorine and fluorine following reactions 4 and 5. The employed rate
 coefficients for reactions 3, 4 and 5 have been determined by
 Hoyermann \& Nacke (1996) and Wallington et al. (1988) at 298 K.
 Reactions 3 to 5, however, do not have an entrance barrier and we can
 retain their value also at much lower temperatures (as a matter
 of fact, a moderate increase of the rate coefficient with decreasing
 temperature is expected in this case). These reactions, well
 characterised in laboratory experiments, were not included in
 previous astrochemical models. We emphasise that they provide a
 direct link between DME and MF, being the former a
 precursor of the latter.

 In addition, the standard loss mechanisms of COMs present in OSU
 (such as the reactions with H$^+$, He$^+$ etc) have all been retained
 and we have added similar loss pathways (with similar rate
 coefficients) also for the new species, such as CH$_3$OCH$_2$, which
 are not present in OSU. In addition, we considered the standard
 reactions of atomic F (Neufeld et al. 2005) and employed the recent
 determination of the rate coefficients for F + H$_2$ by Tizniti et
 al. (2014). As for the reactions of atomic Cl, we note that the Cl +
 H$_2$ reaction is endothermic (Balucani et al. 2004, and references
 therein) and cannot occur in the conditions of the ISM (Neufeld et
 al. 2009). Other reactions between chlorine and small saturated
 molecules are known to be very inefficient. Therefore, Cl is left to
 react with larger molecules, such as those of interest here. Please
 note that the Cl role in the COMs chemistry has been overlooked so
 far, although it can (only) react with large molecules.

\section{Results}\label{sec:results}
Figure \ref{abundances} shows the evolution with time of the
abundances of methoxy, DME and MF, plus CH$_3$ and methanol.  
In the standard case, the methoxy, DME and MF gaseous abundances reach
a peak of $\sim 2\times10^{-10}$, $\sim 2\times10^{-11}$ and $\sim
5\times10^{-12}$, respectively, at $\sim 10^5$ yr and then decrease to
lower values because they freeze-out onto the grain mantles.  The peak
occurs at $\sim 10^5$ yr because of the formation of methanol on the
grains, which occurs when CO starts to freeze-out onto the grain
mantles. A fraction of the methanol is then released in the gas phase
and this triggers the reactions of Fig. 1, which form methoxy, DME and
MF.
In the same panel, we also show the predicted abundances obtained
without considering the new reactions of Table 1. As expected, the
predicted abundances of DME and MF are respectively about three and
one order of magnitude lower.

The panel b of Fig. \ref{abundances} shows the impact on the
abundances when the chemical desorption is varied. The increase by a
factor 10 causes a proportional increase of gaseous methanol and
methoxy, but a larger increase of DME and MF abundances, because
methoxy is the bottleneck to the formation of DME. Caused by the
dominance of reactions 4 and 5, the MF abundance increases by the same
factor than DME. Conversely, putting the chemical desorption to
zero kills the Fig. 1 gas phase reactions as standard cosmic-ray
induced desorption is not effective to desorb methanol from the ices
(as GRAINOBLE considers layer and not bulk chemistry in mantles).

The increase of density by a factor of 5, shown in in the panel c of
Fig. \ref{abundances}, causes a shorter time of residence of the
molecules in the gas phase but only slightly affects the absolute
abundance of the species. On the contrary, the increase of the
temperature to 20 K causes a decrease of the methoxy, DME and MF peak
abundances. This is due to the higher abundance of gaseous atomic H
(whose density is about 1 cm$^{-3}$ at 10 K). At 20 K, the density of
atomic H increases by one order of magnitude due to the higher
evaporation rate.

\begin{figure*}
  \includegraphics[width=11.5cm,angle=0]{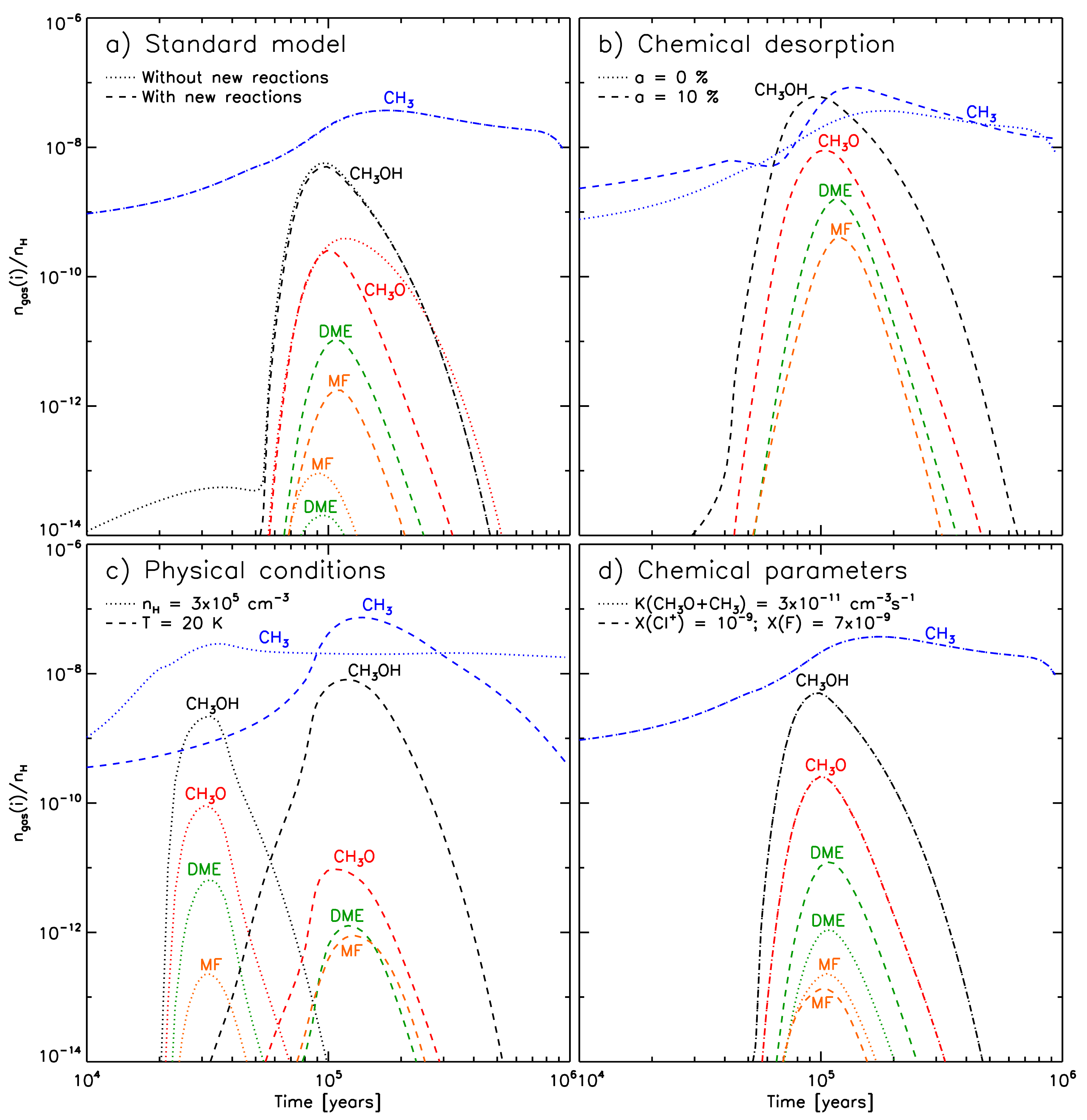}
  \caption{Predicted abundances (with respect to H nuclei) of MF
    (HCOOCH$_3$), DME (CH$_3$OCH$_3$) and methoxy (CH$_3$O) as a
    function of the time, plus some species that intervene in their
    formation, CH$_3$ and methanol (CH$_3$OH). The upper left panel
    (a) refers to the standard case (dashed lines) and the case
    without the new reactions of Table 1 (dotted lines). The upper
    right panel (b) is the same than (a) but with a chemical
    desorption rate equal to 0 (dotted lines) and 10\% (dashed lines)
    respectively. The lower left panel (c) shows a gas with a H
    density of $3\times10^5$ cm$^{-3}$ (dotted lines) and temperature
    20 K (dashed lines). The lower right panel (d) shows the standard
    case but with the rate of reaction 2 of Table 1 10 times lower
    (dotted lines) and with low (see text) Cl and F abundances (dashed
    lines).}
  \label{abundances}      
\end{figure*}
The decrease by a factor 10 of the rate coefficient of reaction 2
leads to a similar decrease of the DME abundance. It also leads to the
MF abundance forming earlier (because of other reactions than reaction
3 of Table 1) than DME and reaching a similar low abundance.  Finally,
decreasing the Cl and F elemental abundance (to the low values quoted
by Graedel et al. 1982) results in a similar behavior of the MF
abundance, as reactions 4 and 5 of Table 1 do not form enough
CH$_3$OCH$_2$.
Note that F is quickly ($\sim10^3$ yr) locked into HF, while Cl
remains mostly atomic until the freeze-out kicks on ($\sim 10^5$
yr). Therefore, reaction 5 of Table 1 plays a major role than reaction
4 in the formation of CH$_3$OCH$_2$ and, consequently, in the
formation of MF from DME.

\section{Discussion}\label{sec:disc}
Whether the proposed new model reproduces the COMs abundances observed
in cold objects depends on the interpretation of where their emission
comes from. Vastel et al. (2014) and Bizzocchi et al. (2014) showed
that methanol emission originates in a shell at the border of the
condensation of the pre-stellar core L1544. These authors claim that
methanol is non-thermally desorbed from the grain mantles and that,
given the lower gas density, the rate of desorption overrun the rate
of freezing-out. Since the same argument in principle applies also to
the COMs, Vastel et al. (2014) suggested that COM emission originates
from the same outer shell. If this is true, the derived abundance of
the COMs detected by Vastel et al. (2014) would be 20 times larger
than if they originated in the cold and dense inner part of the
L1544. Evidently, a similar factor may affect the COMs abundance
derived in the other cold regions (Bacmann et al. 2012; Cernicharo et
al. 2012).

Vastel et al. (2014) estimated a gaseous methanol abundance in the
FUV-illuminated shell of $\sim 6\times10^{-9}$, and an upper limit to
the methoxy, DME and MF abundances of $\sim 2\times10^{-10}$, $\sim
2\times10^{-10}$ and $\sim 2\times10^{-9}$, respectively. These
numbers agree very well with the predictions of our standard model.
On the contrary, the 10\% chemical desorption model predicts too large
abundances. Unfortunately, the upper limit on the MF does not allow to
put constraint on the role of Cl and F, nor on the value of the
reaction 2 rate coefficient. Although we cannot directly compare our predictions
with the absolute abundances quoted by Bacmann et al. (2012) and
Cernicharo et al. (2012) because the assumed H$_2$ column densities
may be overestimated, the DME over MF abundance ratios are compatible
with our predictions. On the contrary, the methoxy abundance measured
by Cernicharo et al. (2012) is 5 times lower than the DME and MF
abundances, at odds with our predictions.

Constraints on the reactions 3 to 5 come from measures of the DME and
MF abundances in different objects. Specifically, Jaber et al. (2014)
showed that there is a linear correlation between these two quantities
over a six orders of magnitude range, confirming a direct link between
DME and MF. The reactions 3 to 5 that we propose here explain in a
natural way that link.
Similarly, we predict the methoxy to be correlated with the methanol
abundance, as it is the product of the reaction of CH$_3$OH with
OH. This reaction has been recently studied and found to be much more
efficient than previously expected (Shannon et al. 2013).
Finally, we predict a possible correlation between methoxy and DME,
although the correlation could be broken if CH$_3$ becomes the
bottleneck of the reaction 2. This correlation is based on the
associative reaction between CH$_3$ and CH$_3$O, a likely process at
10 K, which, however, has never been investigated (\S
\ref{sec:chem-netw-model}). Since the determination of the product
yield and rate coefficients for this reaction will be extremely
difficult, if not impossible, in laboratory experiments, we urge our
colleagues to carry out theoretical calculations of the same kind as
those carried out by Vuitton et al. (2012) for other species.

\section{Conclusions}\label{sec:disc-concl}
We propose a new model to explain the formation of MF, DME and
methoxy, based on recent experimental and theoretical works.
The major conclusions of this work are:\\
i) the non-thermal desorption of iced methanol starts a series of gas
phase reactions leading to the formation of methoxy, DME and MF;\\
ii) Methoxy is formed via the methanol-hydroxyl reaction;\\
iii) DME is mainly formed via the radiative association reaction of
methoxy with CH$_3$;\\
iv) MF is formed from DME via the oxidation of CH$_3$OCH$_2$, a
reaction overlooked by previously models;\\
v) MF is, therefore, the daughter of DME, which explains the observed
correlation (e.g. Jaber et al. 2014).

In conclusion, grain surface chemistry certainly plays a role, for
example in forming hydrogenated species during the prestellar phase
(Tielens \& Hagens 1980; Caselli \& Ceccarelli 2012), but not
necessarily in the formation of all COMs. Gas phase and grain surface
chemistry will have to share the reign of the formation of COMs.
 
\section*{Acknowledgments}
NB acknowledges the financial support from the Universit\'e Joseph
Fourier and the Observatoire de Grenoble, CC from the French Space
Agency CNES, VT from the NASA postdoctoral program. We thank
S.J. Klippenstein for useful discussions on radiative association
reactions, and A. Jaber and C. Vastel for exchanges on their work. We
also thank an anonymous referee whose comments helped to improve the
article.

{}

\bsp

\label{lastpage}


\begin{thebibliography}{}

\bibitem[Bacmann et~al.2012)]{ba12}
{Bacmann}, A., {Taquet}, V., {Faure}, A., {Kahane}, C., \& {Ceccarelli}, C.
 2012, A\&A, 541, L12

\bibitem[Balucani2004]{bal04}
Balucani, N., Skouteris, D., Capozza, G. et al. 2004, Phys. Chem. Chem. Phys. 6, 5007

\bibitem[biz2014]{bi14}
Bizzocchi L., Caselli P., Spezzano S., Leonardo E., 2014, A\&A
569, 27

\bibitem[{{Blake} {et~al.}(1987){Blake}, {Sutton}, {Masson}, \&
 {Phillips}}]{bl87}
{Blake}, G.~A., {Sutton}, E.~C., {Masson}, C.~R., {Phillips}, T.~G. 1987,
ApJ 315, 621

\bibitem[Bottinelli et al.(2007)]{Bottinelli2007}
{Bottinelli}, S., {Ceccarelli}, C., {Williams}, J.~P., {Lefloch}, B. 2007,  A\&A 463, 601-610

\bibitem[{{Caselli} {et~al.}(1993){Caselli}, {Hasegawa}, \& {Herbst}}]{ca93}
{Caselli}, P., {Hasegawa}, T.~I., {Herbst}, E. 1993, ApJ 408, 548

\bibitem[ca2012]{ca12}
Caselli, P., Keto, E., Bergin, E.A., et al. 2012 ApJL 759, L37

\bibitem[{{Caselli} \& {Ceccarelli}(2012)}]{ca12}
{Caselli}, P., \& {Ceccarelli}, C. 2012,  A\&A Rev, 20, 56

\bibitem[{{Cazaux} {et~al.}(2003){Cazaux}, {Tielens}, \& {Ceccarelli}}]{ca03}
{Cazaux}, S., {Tielens}, A.~G.~G.~M., {Ceccarelli}, C. et al. 2003, ApJL
 593, L51


\bibitem[{{Cernicharo} {et~al.}(2012){Cernicharo}, {Marcelino}, \&
 {Roueff}}]{ce12}
{Cernicharo} J., {Marcelino} N., {Roueff} E. et al. 2012, ApJL 759, L43

\bibitem[changherbst]{ch2014}
Chang, Q. \& Herbst, E., 2014, ApJ 787, 135

\bibitem[{{Charnley} {et~al.}(1992){Charnley}, {Tielens}, \& {Millar}}]{ch92}
{Charnley}, S.~B., {Tielens}, A.~G.~G.~M., {Millar}, T.~J. 1992, ApJL 399,
 L71

\bibitem[dulieu2013]{du13}
Dulieu, F., Congiu, E., Noble, J., et al. 2013, NatSR 3, 1338

\bibitem[GarrodHerbst]{gh06}
{Garrod}, R.~T. \& {Herbst}, E., 2006, A\&A 457, 927

\bibitem[Garrod2007]{ga07}
{Garrod}, R.~T., {Wakelam}, V., {Herbst}, E., 2007, A\&A 467, 1103

\bibitem[{{Geppert} {et~al.}(2007){Geppert}, {Vigren}, \& {Hamberg}}]{ge07}
{Geppert}, W.~D., {Vigren}, E., \& {Hamberg}, M. et al.2007, in European
 Planetary Science Congress 2007, 613

\bibitem[grae2]{gr82}
Graedel, T. E., Langer, W. D., \& Frerking, M. A. 1982, ApJS, 48, 321

\bibitem[has92]{h92}
Hasegawa, T. I., Herbst, E., M., L. C. 1992, ApJ Supp. 82, 167

\bibitem[has93]{h93}
Hasegawa, T. I. \& Herbst, E. 1993, MNRAS 261, 83

\bibitem[{{Horn} {et~al.}(2004){Horn}, {M{\o}llendal}, \& {Sekiguchi}}]{ho04}
{Horn}, A., {M{\o}llendal}, H., {Sekiguchi}, O. et al. 2004, ApJ, 611, 605

\bibitem[Hoyermann]{ho}
Hoyermann K. \& Nacke F. 1996, Symp.Int.Combust.Proc. 26, 505

\bibitem[Jaber et al.(2014)]{jaber} 
 {Jaber}, A., {Ceccarelli}, C., {Kahane}, C., {Caux}, E. 2014, ApJ 791, 29

\bibitem[{{Millar} {et~al.}(1991){Millar}, {Herbst}, \& {Charnley}}]{mi91}
{Millar}, T.~J., {Herbst}, E., {Charnley}, S.~B. 1991, ApJ, 369,
147

\bibitem[hh08]{hh08}
Harada, N. \& Herbst, E. 2008, ApJ 685, 272

\bibitem[md14]{md14}
Minissale, M. \& Dulieu, F. 2014, J.Chem.Phys. 141, 14304

\bibitem[neufeld05]{neufeld2005}
Neufeld, D., Wolfire, M., Schilke, P. 2005, ApJ 628, 260

\bibitem[neufeld09]{neufeld2009}
Neufeld, D. \& Wolfire, M. 2009, ApJ 706, 1594

\bibitem[{{{\"O}berg} {et~al.}(2010){{\"O}berg}, {Bottinelli}, {J{\o}rgensen},
 \& {van Dishoeck}}]{ob10}
{{\"O}berg}, K.~I., {Bottinelli}, S., {J{\o}rgensen}, J.~K., {van Dishoeck},
 E.~F. 2010, ApJ 716, 825

\bibitem[{{Requena-Torres} {et~al.}(2006){Requena-Torres},
    {Martin-Pintado}, {Martin}, \& {Amo-Baladron}}]{re07}
  {Requena-Torres}, M.~A., {Martin-Pintado}, J., 
  {Rodriguez-Franco}, A. et al. 2006,  A\&A 455, 971

\bibitem[rimola14]{ri14}
Rimola, A., Taquet V., Ugliengo, P., Balucani, N., Ceccarelli, C. 2014, A\&A 572, 70

\bibitem[Shannon et al.(2013)]{sh}
Shannon, R. J., Blitz, M. A., Goddard, A., Heard, D. E. 2013, Nature Chemistry 5, 745.

\bibitem[Sivaramakrishnan et al. (2011)] {}
Sivaramakrishnan, R.,  Michael, J. V.,  Wagner, A. F. et al. 2011, 
Combustion and Flame 158, 618.

\bibitem[song]{so}
Song X., Hou H., Wang B. 2005, Phys.Chem.Chem.Phys. 7, 3980

\bibitem[{{Taquet} {et~al.}(2012){Taquet}, {Ceccarelli}, \& {Kahane}}]{ta12}
{Taquet}, V., {Ceccarelli}, C., {Kahane}, C. 2012,  A\&A 538, A42

\bibitem[Taquet13]{t13}
Taquet, V., Peters, P.S., Kahane, C., et al. 2013, A\&A 550, 127

\bibitem[Taquet14]{t14}
Taquet, V., Charnley, S.B., Sipila, O. 2014, ApJ 791, 1

\bibitem[{{Tielens} \& {Hagen}(1982)}]{ti82}
{Tielens}, A.~G.~G.~M., \& {Hagen}, W. 1982,  A\&A 114, 245

\bibitem[Tizniti]{tiz}
Tizniti M., Le Picard S.D., Lique F. et al. 2014, Nat.Chem.  6, 141

\bibitem[vastel2014]{v14}
Vastel C., Ceccarelli C., Lefloch B., Bachiller R. 2014, ApJL 795, L2

\bibitem[{{Vasyunin} \& {Herbst}(2013)}]{va13}
{Vasyunin}, A.~I., \& {Herbst}, E. 2013, ApJ 769, 34, VH2013

\bibitem[Vuitton et al. (2012)]{vu12}
Vuitton, V., Yelle, R. V.,  Lavvas, P., Klippenstein, S. J. 2012, ApJ 744, 11 

\bibitem[Wake2008]{wa2008}
Wakelam, V. \& Herbst, E. 2008, Ap.J 680, 371

\bibitem[Wallinton et al. 1988]{wa88}
Wallington T.J., Skewes L.M., Siegel W.O., Wu C-H., Japar S.M. 1988, Int.J.Chem.Kinet. 20, 867

\end{thebibliography}
\end{document}